\documentclass[aps,prd,nofootinbib,preprint,tightenlines,showpacs,amsmath,amssymb]{revtex4}
\newcommand{\be}{\begin{equation}}
\newcommand{\ee}{\end{equation}}
\newcommand{\bea}{\begin{eqnarray}}
\newcommand{\eea}{\end{eqnarray}}
\def\bml{\begin{subequations}}
\def\blea{\bml\begin{eqnarray}}
\def\elea{\end{eqnarray}\end{subequations}}
\def\bE{\mathbf{E}}
\def\bX{\mathbf{X}}
\def\Rmax{R_{\text{max}}}
\def\dRmax{R'_{\text{max}}}
\def\ddRmax{R''_{\text{max}}}
\def\Texotic{T_{\text{exotic}}}
\def\Gexotic{G_{\text{exotic}}}
\def\lpl{l_{\text{Planck}}}
\def\bl{\boldsymbol{\ell}}
\def\bk{\mathbf{k}}
\def\bv{\mathbf{v}}
\def\bx{\mathbf{x}}
\def\by{\mathbf{y}}

\usepackage{epsfig}

\begin{document}

\title{Averaged null energy condition in a classical curved background}

\author{Eleni-Alexandra Kontou}
\author{Ken D. Olum}
\affiliation{Institute of Cosmology, Department of Physics and Astronomy,\\ 
Tufts University, Medford, MA 02155, USA}

\begin{abstract}

The Averaged Null Energy Condition (ANEC) states that the integral
along a complete null geodesic of the projection of the stress-energy
tensor onto the tangent vector to the geodesic cannot be negative.
Exotic spacetimes, such as those allow wormholes or the construction
of time machines are possible in general relativity only if ANEC is
violated along achronal geodesics.  Starting from a conjecture that
flat-space quantum inequalities apply with small corrections in
spacetimes with small curvature, we prove that ANEC is obeyed by a
minimally-coupled, free quantum scalar field on any achronal null
geodesic surrounded by a tubular neighborhood whose curvature
is produced by a classical source.

\end{abstract}


\maketitle

\section{Introduction}

It is always possible to invent a spacetime with exotic features, such
as wormholes, superluminal travel, or the construction of time
machines, and then determine what stress-energy tensor is necessary to
support the given spacetime.  To rule out such exotic spacetimes we
would like to prove energy conditions that restrict the stress-energy
tensor that might arise from quantum fields and show that the
stress-energy necessary to support an exotic spacetime is impossible.
We need a condition which is strong enough to rule out exotic cases
while simultaneously weak enough to be proven correct, or at least to
be free of known counterexamples.

The best possibility for such a condition seems to be the achronal
averaged null energy condition \cite{Graham:2007va}, which requires
the following.  Let $M$ be a manifold with Lorentzian metric $g$ and
$T$ be the stress-energy tensor of some fields on $M$.  Let
$\gamma$ be a complete null geodesic with tangent vector $\bl$.
Suppose that $\gamma$ is achronal, i.e., no two points of $\gamma$ can
be connected by a timelike curve.  Then
\be
\int_\gamma T_{ab} \ell^a \ell^b \ge 0\,.
\ee
That is to say, we require that the projection of the stress-energy
tensor along a geodesic integrate to a non-negative value, but only
for geodesics that are achronal.  As far as we know there are no known
violations of achronal ANEC using minimally-coupled scalar
fields.\footnote{Non-minimally coupled scalar fields have some unique
  properties, which we discuss briefly in Sec.~\ref{sec:discussion}.}
Achronal ANEC is sufficient to rule out many exotic spacetimes
\cite{Graham:2007va}.

Reference \cite{Fewster:2006uf} proved that the averaged null energy
condition (ANEC) holds for geodesics traveling through empty, flat
space, even if elsewhere in the spacetime there are boundaries or
spacetime curvature, providing that these stay some minimum distance
from the geodesic and do not affect the causal structure of the
spacetime near the geodesic.  Here we will extend this work to
geodesics traveling in curved spacetime, with the restriction that the
spacetime near the geodesic must obey the null convergence condition,
\be\label{eqn:nullconvergence}
R_{ab}V^a V^b\ge0
\ee
for any null vector $V^a$.  Equation~(\ref{eqn:nullconvergence})
holds whenever the curvature is generated by a ``classical
background'' whose stress tensor obeys the null energy condition
(NEC),
\be\label{eqn:NEC}
T_{ab}V^a V^b\ge0\,.
\ee
`We stress that Eqs.~(\ref{eqn:nullconvergence}) and (\ref{eqn:NEC})
need not hold in general, but only in a neighborhood of the null
geodesic on which we seek to prove ANEC.  Thus, for example, the
results of this paper apply to any geodesic which does not encounter
any material source, even if such sources exist elsewhere in the
spacetime.

Reference~\cite{Fewster:2006uf} used a null-contracted
timelike-averaged quantum inequality proved for flat space in
Ref.~\cite{Fewsterromanwitherratum}.  Here we will conjecture that this
quantum inequality holds with a small modification in spacetimes with
small curvature.  We will then be able to rule out ANEC violation,
subject to several conditions.

In the next section we give the conditions on which our theorem
depends.  In Sec.~\ref{sec:theorem} we state our theorem.  In
Sec.~\ref{sec:QI} we discuss what it means to have small curvature and
state our conjecture.  In Sec.~\ref{sec:proof} we prove the theorem,
and in Sec.~\ref{sec:discussion} we conclude with a discussion of
remaining possibilities for the generation of exotic spacetimes.  We
use the sign convention $(+,+,+)$ in the classification of Misner,
Thorne and Wheeler \cite{MTW}.

\section{Assumptions}
\subsection{Congruence of geodesics}\label{sec:classical}

As in Ref.~\cite{Fewster:2006uf}, we will not be able to rule out ANEC
violation on a single geodesic.  However, a single geodesic would
not lead to an exotic spacetime.  It would be necessary to have ANEC
violation along a finite congruence of geodesics in order to have a
physical effect.

So let us suppose that our spacetime contains a null geodesic $\gamma$
with tangent vector $\bl$ and that there is a ``tubular neighborhood''
$M'$ of $\gamma$ composed of a congruence of achronal null geodesics,
defined as follows.  Let $p$ be a point of $\gamma$, and let $M_p$ be
a normal neighborhood of $p$.  Let $\bv$ be a null vector at $p$,
linearly independent of $\bl$, and let $\bx$ and $\by$ be spacelike
vectors perpendicular to $\bv$ and $\bl$.  Let $q$ be any point in
$M_p$ such that $p$ can be connected to $q$ by a geodesic whose
tangent vector is in the span of $\{\bv,\bx,\by\}$.  Let $\gamma(q)$
be the geodesic through $q$ whose tangent vector is the vector $\bl$
parallel transported from $p$ to $q$.  If a neighborhood $M'$ of
$\gamma$ is composed of all geodesics $\gamma(q)$ for some choice of
$p$, $M_p$, $\bv$, $\bx$ and $\by$, we will say that $M'$ is a tubular
neighborhood of $\gamma$.

\subsection{Coordinate system}\label{sec:coordinates}
Given the above construction, we can define Fermi-like coordinates
\cite{Kontou:2012kx} on $M'$ as follows.  Without loss of generality
we can take the vector $\bv$ to be normalized so that $\bv \cdot \bl
=-1$, and $\bx$ and $\by$ to be unit vectors.  Then we have a
pseudo-orthonormal tetrad at $p$ given by $\bE_{(u)} = \bl$,
$\bE_{(v)} = \bv$, $\bE_{(x)} = \bx$, and $\bE_{(y)} = \by$.  The
point $q = (u,v,x,y)$ in these coordinates is found as follows.  Let
$q^{(1)}$ be found by traveling unit affine parameter from $p$ along the
geodesic generated by $v \bE_{(v)} + x \bE_{(x)}+y \bE_{(y)}$.  Then
$q$ is found by traveling unit affine parameter from $q^{(1)}$ along the
geodesic $u \bE_{(u)}$.  During this process the tetrad is parallel
transported.  All vectors and tensors will be described using this
transported tetrad unless otherwise specified.  We will use Latin
letters from the beginning of the alphabet to denote arbitrary
components in the tetrad basis.

The points with $u$ varying but other coordinates fixed form one of
the null geodesics of the previous section.

\subsection{Curvature}
\label{sec:curvature}
We suppose that the curvature inside $M'$ obeys the null convergence
condition, Eq.~(\ref{eqn:nullconvergence}).  We will refer to this as
a ``classical background'', but the only way it need be classical is
Eq.~(\ref{eqn:nullconvergence}).

We would not expect any energy conditions to hold when the curvature
is arbitrarily large, because then we would be in the regime of
quantum gravity, so we will require that the curvature be bounded.
In the coordinate system of Sec.~\ref{sec:coordinates} we require
\be\label{eqn:Rbound}
|R_{abcd}|<\Rmax\,,
\ee
everywhere in $M'$.

We will also need to bound the first and second derivatives of the
Riemann tensor,
\be
|R_{abcd,e}|<\dRmax, \qquad |R_{abcd,ef}|<\ddRmax
\ee
everywhere in $M'$.  The bounds $\Rmax$, $\dRmax$ and $\ddRmax$
are some (independent) finite numbers, but they need not be small.

We will also assume that the curvature is smooth.

\subsection{Causal structure}

We will also require that conditions outside $M'$ do not affect the
causal structure of the spacetime in $M'$
\cite{Fewster:2006uf}\footnote{This condition is equivalent to 
$J^-(p,M) \cap M'=J^-(p,M')$ for all $p \in M'$.}
\be\label{eqn:causal}
J^+(p,M) \cap M'=J^+(p,M')
\ee
for all $p \in M'$. 
Otherwise the curvature outside $M'$ may be
arbitrary.

\subsection{Quantum field theory} \label{sec:quantum}

We consider a quantum scalar field in $M$.  We will work entirely
inside $M'$, and there we require that the field be free and minimally
coupled.  It may be massive or massless.  Outside $M'$, however, we
can allow different curvature coupling, interactions with other
fields, and even boundary surfaces with specified boundary conditions.

Because $M$ may not be globally hyperbolic, it is not completely
straightforward to specify what we mean by a quantum field theory on
$M$.  We will use the same strategy as Ref.~\cite{Fewster:2006uf}.
Our results will hold for any quantum field theory on $M$ that reduces
to the usual quantum field theory on each globally hyperbolic
subspacetime of $M$.  The states of interest will be those that
reduce to Hadamard states on each globally hyperbolic subspacetime,
and we will refer to any such state as ``Hadamard''.  See Sec.~II~B
of Ref.~\cite{Fewster:2006uf} for further details.

\section{The theorem} \label{sec:theorem}
We can now state our theorem.

\emph{Theorem 1.}  Let $(M,g)$ be a (time-oriented) spacetime and let
$\gamma$ be a null geodesic on $(M,g)$, and suppose that $\gamma$ is
surrounded by a tubular neighborhood $M'$ in the sense of
Sec.~\ref{sec:classical}, obeying the null convergence condition,
Eq.~(\ref{eqn:nullconvergence}), and that we have constructed
coordinates by the procedure of Sec.~\ref{sec:coordinates}.  Suppose
that the curvature in this coordinate system is smooth and obeys the
bounds of Sec.~\ref{sec:curvature}, that the curvature in the system
is localized, i.e., in the distant past and future the spacetime is
flat, and that the causal structure of $M'$ is not affected by
conditions elsewhere in $M$, Eq.~(\ref{eqn:causal}).

Let $\omega$ be a state of the free minimally coupled quantum scalar
field on $M'$ obeying the conditions of Sec.~\ref{sec:quantum}, and
let $T$ be the renormalized expectation value of the
stress-energy tensor in state $\omega$.

\emph{Under these conditions, it is impossible for the ANEC integral,
\be
A=\int_{-\infty}^{\infty} d \lambda\, T_{ab}\ell^a \ell^b (\Gamma(\lambda)),
\ee
to converge uniformly to negative values on all geodesics
$\Gamma(\lambda)$ in $M'$.}

In the next section, we will conjecture that a known flat-space
quantum inequality can be extended to spacetimes with small curvature
in a particular way.  From this conjecture we will be able to prove
Theorem 1.

\section{Quantum Inequality}\label{sec:QI}

The proof will proceed very much along the lines of
Ref.~\cite{Fewster:2006uf}.  That paper used the following quantum
inequality for the null-projected but timelike-averaged stress-energy
tensor, derived by Fewster and Roman
\cite{Fewster:1999gj,Fewsterromanwitherratum}. Let $w(\tau)$ be a
timelike geodesic segment parameterized by proper time
$\tau\in(-\tau_0,\tau_0)$.  Let $g(\tau)$ be a smooth real function
with compact support contained in $(-\tau_0,\tau_0)$.  Let $\bk$ be
the tangent vector to $w(\tau)$ and let $\bl$ be a constant null
vector.  Let $T$ be the renormalized stress-energy tensor of a
massless or massive\footnote{The derivation of
  \cite{Fewsterromanwitherratum} was for the massless case, but the
  same argument holds in the massive case as well
  \cite{Fewster:2006uf}.}  minimally-coupled quantum scalar field in a
Hadamard state.  Then the projection of $T$ on the null vector $\bl$
obeys a quantum inequality when integrated along the timelike geodesic
$w$,
\be\label{eqn:QNEI1}
\int_{-\tau_0}^{\tau_0}d\tau\, T_{ab}(w(\tau))\ell^a \ell^b g(\tau)^2 \geq
-\frac {(k_a \ell^a)^2}{12 \pi^2}\int_{-\tau_0}^{\tau_0}d\tau
g''(\tau)^2\,.
\ee
Equation~(\ref{eqn:QNEI1}) is a consequence of the result of
Ref.~\cite{Fewster:1999gj}, which applies to general worldlines in
curved spacetime.  This more general result is in the form of a
``difference inequality'' that restricts the amount by which the
left-hand side of Eq.~(\ref{eqn:QNEI1}) can be more negative than the
same quantity evaluated in a reference state.  We need an absolute
bound, such as Eq.~(\ref{eqn:QNEI1}), but applicable to curved
spacetime.  While such a bound has not been proven, we conjecture that
Eq.~(\ref{eqn:QNEI1}) can be extended to spacetimes of small
curvature.

The basic idea was given by Ford and Roman \cite{Ford:1995wg}.
Suppose that we want to test Eq.~(\ref{eqn:QNEI1}) in a laboratory on
the surface of the earth.  We are not in flat space, but rather in
space with curvature of order $G M_\oplus/R_\oplus^3$.  Furthermore
the apparatus for measuring $T$ might not be in free fall but rather
accelerating with the acceleration due to gravity at the earth's
surface, $a = G M_\oplus/R_\oplus^2$.  But in a laboratory-scale
experiment, these differences should not matter.  We expect
Eq.~(\ref{eqn:QNEI1}) to hold with a small correction for almost
geodesic $w(\tau)$ in spacetimes with small curvature.

What does it mean for the curvature to be small?  First of all, since
the curvature has dimensions $(\text{length})^{-2}$, we have to multiply by the
square of some length to get a number that we can require to be much
less than 1.  The obvious length in the present example is $\tau_0$.

We also face a problem that curvature is a tensor, and we would like
to make coordinate-invariant statements.  In a Riemannian space, we
could require, for example, that the sectional curvature of each plane
in the tangent space at each point be small.  But in a Lorentzian
spacetime this does not work: the sectional curvature is never bounded
unless it is constant \cite{Kulkarni:curvature,Thorpe:curvature}.  A
simple example of the problem is that the spacetime could contain a
plane gravitational wave.  The amplitude of such a wave is entirely
dependent on the reference frame; it can be made arbitrarily small or
arbitrarily large by the choice of coordinates.  Thus one cannot say
that all components of the Riemann tensor are small without regard to
coordinate system.

Fortunately, in our case, we have a privileged observer whose
stress-energy tensor we want to integrate.  Thus the worldline of
that observer can be used to generate a preferred coordinate
system.\footnote{A similar technique was used in
  Ref.~\cite{Ford:1995wg}.}  This works straightforwardly on that
worldline, but to apply this idea to other places in the spacetime we
will have to parallel transport the observer's 4-velocity.
Fortunately, in the case where the curvature is in fact small, the
precise details of this transport will not matter.

With these considerations in mind we proceed as follows.  Let $(N,g)$
be a globally hyperbolic spacetime and let $w(\tau)$ be a timelike
path in $N$, parameterized by proper time $\tau\in(-\tau_0,\tau_0)$,
with tangent vector $\bk$.  In general we will only need to consider the
``double cone'' $N = J_-(w(\tau_0))\cap J_+(w(-\tau_0))$.  Let
$\epsilon \ll 1$.  We will say that $(N,g)$ has small
curvature $\epsilon $ relative to $w$ if $N$ is a normal neighborhood
of the point $p = w(0)$ and there exists a set of three unit spacelike
vectors $\bE_{(i)}$, $i=1,2,3$ at $p$, orthogonal to each other and to
$\bE_{(0)} = k$, such that at each point $q$, every component of the
Riemann tensor in the tetrad basis formed by parallel transporting the
tetrad $\{\bE_{(a)}\}$ along the geodesic connecting $p$ and $q$
obeys
\be\label{eqn:smallcurvature}
|R_{abcd}| \tau_0^2 < \epsilon\,.
\ee

Suppose $(N,g)$ has small curvature $\epsilon$ by the above
definition, and we consider the curvature components in a different
tetrad basis resulting from a choice of $\bE_{(i)}$ other than the one
which satisfies Eq.~(\ref{eqn:smallcurvature}).  Changing to such a
basis will given curvature components that are linear combinations of
the ones we had before, and so may be larger than the bound of
Eq.~(\ref{eqn:smallcurvature}), but only by factors of order 1.

We could also choose a different starting point $p$ on $w$.  Since the
curvature is small, the different parallel transport would change the
Riemann tensor components only by factors of $1+O(\epsilon)$, so the
condition would be the same at first order.

We will also require that the proper acceleration of the path on which
we want the quantum inequality to hold should be small.  Since
acceleration has the units of inverse time, we will multiply by the
time $\tau_0$ to get a dimensionless measure limiting the total
acceleration along the path of interest.

Once we are in curved spacetime, we must address ambiguities in the
definition of the stress-energy tensor $T$.  We will adopt the
axiomatic definition given by Wald \cite{Wald:qft}, but there remains
the ambiguity of adding local curvature terms with arbitrary
coefficients.  These terms are the metric, the Einstein tensor, and
two terms that are second order in the curvature or involve second
derivatives of the curvature \cite{Birrellbook},
\bml\label{eqn:12H}
\bea
^{(1)}H_{ab} &=& 2R_{;ab} -2g_{ab}\Box R + g_{ab}R^2/2 - 2RR_{ab}\\
^{(2)}H_{ab} &=& R_{;ab} -\Box R_{ab} -g_{ab}\Box R/2
+g_{ab}R^{cd} R_{cd}/2  - 2R^{cd}R_{acbd} \,.
\elea
A multiple of the the metric will not concern us here, because it
vanishes when contracted with the null vector $\bl$.  A term
proportional to the Einstein tensor can be absorbed into
renormalization of Newton's constant, and we assume that that has been
done.

As it turns out, the remaining ambiguity will not affect our proof
below.  However, it must be taken into account in the present
conjecture.  Following an idea in Ref.~\cite{Fewster:2007rh}, we will
allow any definition of $T_{ab}$ and absorb the ambiguity into a
local curvature term in our bound.

We now can now conjecture that Eq.~(\ref{eqn:QNEI1}) holds with a
modification of order $\epsilon$ and a local curvature term.

\emph{Conjecture 1.}  Let $(N,g)$ be a globally hyperbolic spacetime
and let $w(\tau)$ be a timelike path in $N$, parameterized by proper
time $\tau\in(-\tau_0,\tau_0)$.  Let $\bk$ be the tangent vector to
$w$ and let $\bl$ be a null vector field obeying $k^a \nabla_a \ell^b
= 0$.  Let $g(\tau)$ be a smooth real function with compact support
contained in $(-\tau_0,\tau_0)$.  Let $T$ be any definition (obeying
Wald's axioms \cite{Wald:qft}) of the renormalized stress-energy
tensor of a massless or massive minimally-coupled quantum scalar field
in a Hadamard state.  If $(N,g)$ has small curvature $\epsilon$
relative to $w$ and $|D^2w^a/d\tau^2|\tau_0 < \epsilon$ everywhere on
$w$, then
\be\label{eqn:QNEI2}
\int_{-\tau_0}^{\tau_0}d\tau\, T_{ab}(w(\tau))\ell^a \ell^b g(\tau)^2 \geq
-\frac {(k_a \ell^a)^2}{12 \pi^2}\int_{-\tau_0}^{\tau_0}d\tau\,
g''(\tau)^2[1+c(\epsilon)]
+\int_{-\tau_0}^{\tau_0}d\tau\, g(\tau)^2 C_{ab}\ell^a \ell^b
\,,
\ee
where $c(\epsilon)$ is a function that goes to zero as $\epsilon\to0$,
and $C_{ab}$ is a linear combination of Eqs.~(\ref{eqn:12H}).  The
form of $c(\epsilon)$ and the coefficients of $^{(1)}H$ and $^{(2)}H$
in $C_{ab}$ do not depend on the spacetime or the quantum state.  Note
that terms in Eqs.~(\ref{eqn:12H}) whose tensor structure is that of
the metric do not contribute in Eq.~(\ref{eqn:QNEI2}) because $\bl$ is
null.

We intend to prove Conjecture 1 in future work.

\section{Proof of the Theorem}\label{sec:proof}

\subsection{Outline of the proof}

Following Ref.~\cite{Fewster:2006uf}, we will prove Theorem 1 by
contradiction using integrals over a parallelogram shown below in
Fig.~\ref{fig:parallelogram}.  By considering this parallelogram as
made up of segments of the null geodesics of $M'$, and assuming
Theorem 1 is violated, we set a negative upper bound on the integral
of the null-contracted stress-energy tensor over the parallelogram.
Then we consider the same set of points as being made up of timelike
paths, and demonstrate that these paths obey the conditions of
Conjecture 1.  Thus using Eq.~(\ref{eqn:QNEI2}), we can set a lower
bound on the same integral over the parallelogram.  In the limit where
the parallelogram becomes long and narrow, these bounds conflict,
proving the theorem.

\subsection{The parallelogram}
We will use the $(u,v,x,y)$ coordinates of
Sec.~\ref{sec:coordinates}. Let $r$ be a positive number small enough such that
whenever $|v|,|x|,|y|<r$, the point $(0,v,x,y)$ is inside the
normal neighborhood $N_p$ defined in Sec.~\ref{sec:classical}.  Then
the point $(u,v,x,y) \in M'$ for any $u$.

\begin{figure}
\epsfysize=70mm
\epsfbox{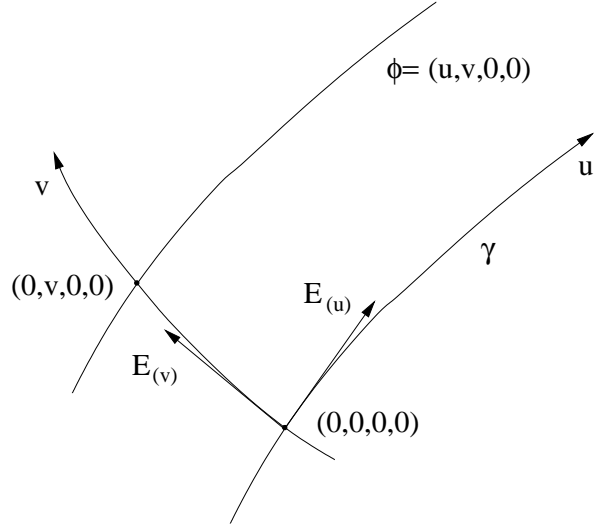}
\caption{Construction of the family of null geodesics $\Phi$ using Fermi normal coordinates}
\label{fig:fermi}
\end{figure}

Now consider the points
\be\label{eqn:Phi}
\Phi(u,v) = (u,v,0,0)\,.
\ee
With $v$ fixed and $u$ varying, these are null geodesics in $M'$.
(See Fig.~\ref{fig:fermi}.)
Write the ANEC integral
\be\label{eqn:Av}
A(v) = \int_{-\infty}^\infty du\, T_{uu}(\Phi(u,v))\,.
\ee
Suppose that, contrary to Theorem 1, Eq.~(\ref{eqn:Av}) converges uniformly to
negative values for all $|v| < r$.  We will prove that this leads to a
contradiction.

Since the convergence is uniform, $A(v)$ is continuous.  Then 
since $A(v)<0$ for all $|v| < r$, we can choose a positive number $v_0 < r$
and a negative number $-A$ larger than all $A(v)$ with $v \in
(-v_0, v_0)$.  Then it is possible to find some number $u_1$ large enough that
\be\label{eqn:uintegral}
\int_{u_-(v)}^{u_+(v)}du\,  T_{uu}(\Phi(u,v)) < -A/2
\ee
for any $v \in (-v_0, v_0)$ as long as \bml \label{eqn:uinequality}
\bea
u_+(v)&>&u_1 \\
u_-(v)&<&-u_1\,.
\elea

As in Ref.~\cite{Fewster:2006uf}, we will define a series of
parallelograms in the $(u, v)$ plane, and derive a contradiction by
integrating over each parallelogram in null and timelike directions.
Each parallelogram will have the form 
\bml\label{eqn:uvrange}
\bea
v &\in& (-v_0, v_0)\\
u &\in& (u_-(v),u_+(v))\,,
\elea
where $u_-(v),u_+(v)$ are
linear functions of $v$ obeying Eqs.~(\ref{eqn:uinequality}).  On each
parallelogram we will construct a weighted integral of
Eq.~(\ref{eqn:uintegral}) as follows.  Let $f(a)$ be a smooth function
supported only within the interval $(-1,1)$ and normalized
\be
\int_{-1}^1 da f(a)^2 = 1\,.
\ee
Then we can write
\be\label{eqn:uvintegral}
\int_{-v_0}^{v_0} dv\, f(v/v_0)^2 \int_{u_-(v)}^{u_+(v)} du\,T_{uu}(\Phi(u,v)) < -v_0A/2\,.
\ee

We can construct this same parallelogram as follows.  First choose
a velocity V.  Eventually we will take the limit $V \to 1$.  Define
the Doppler shift parameter
\be
\delta= \sqrt{\frac{1+V}{1-V}}\,.
\ee
Let $\alpha$ be some fixed number with $0 <\alpha <1/3$ and then let
\be\label{eqn:tau0}
\tau_0 =\delta^{-\alpha} r\,.
\ee
As $V \to 1$, $\delta\to \infty$ and $\tau_0 \to 0$.

\begin{figure}
\epsfysize=70mm
\epsfbox{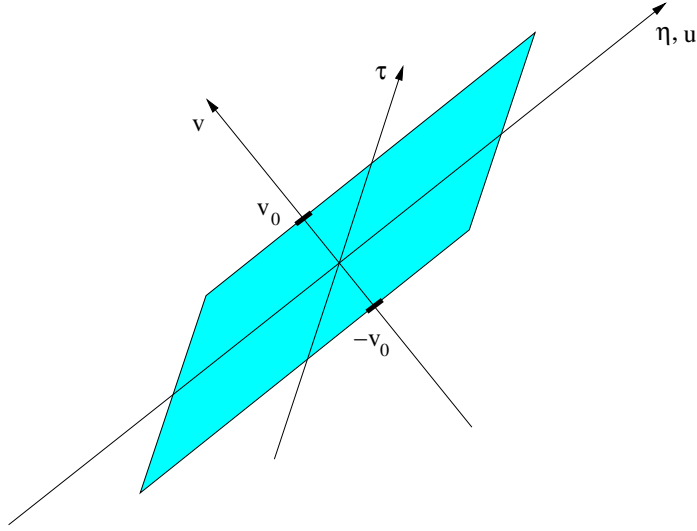}
\caption{The parallelogram $\Phi(u,v)$, $v \in (-v_0, v_0)$,
$u \in (u_-(v),u_+(v))$, or equivalently $\Phi_V(\eta,\tau)$,
  $\tau\in(-\tau_0,\tau_0)$, $\eta \in (-\eta_0,\eta_0)$}
\label{fig:parallelogram}
\end{figure}

Now define the set of points
\be\label{eqn:PhiV}
\Phi_V(\eta,\tau)=\Phi(\eta+\frac{\delta \tau}{\sqrt{2}},
\frac{\tau}{\sqrt{2} \delta})\,.
\ee
We will be interested in the paths given by $\Phi_V(\eta,\tau)$ with
$\eta$ fixed and $\tau$ ranging from $-\tau_0$ to $\tau_0$.  In flat
space, such paths would be timelike geodesic segments, parameterized
by $\tau$ and moving at velocity $V$ with respect to the original
coordinate frame.  In our curved spacetime, this is nearly the case,
as we will show below.
Define 
\blea
\eta_0&=&u_1+\tau_0 \delta/\sqrt{2}\label{eqn:eta0}\\
v_0&=&\tau_0/(\sqrt{2} \delta)\\
u_{\pm}(v)&=&\pm \eta_0+\delta^2 v
\elea
so that $u_{\pm}$ satisfies Eqs.~(\ref{eqn:uinequality}). Then the
range of points given by Eq.~(\ref{eqn:Phi}) with coordinate ranges
specified by Eqs.~(\ref{eqn:uvrange}) is the same as that given by
Eq.~(\ref{eqn:PhiV}) with coordinate ranges
\blea
-\tau_0&<&\tau<\tau_0\\
-\eta_0&<&\eta<\eta_0
\elea
The parallelogram is shown in Fig.~\ref{fig:parallelogram}.

The Jacobian
\be
\left|\frac {\partial (u,v)}{\partial (\eta,\tau)}\right| =
\frac{1}{\sqrt{2}\delta}
\ee
so Eq.~(\ref{eqn:uvintegral}) becomes
\be\label{eqn:etintegral}
\int_{-\eta_0}^{\eta_0} d\eta \int_{-\tau_0}^{\tau_0} d\tau\,
T_{uu}(\Phi_V(\eta,\tau)) f(\tau/\tau_0)^2 < -A\tau_0/2\,.
\ee
We will show that this is impossible by applying the quantum inequality
of Sec.~\ref{sec:QI}.

\subsection{Transformation of the Riemann tensor}

We would like to work in coordinates which bring to rest, as much as
possible, the path $\Phi_V(\eta,\tau)$ with $\eta$ fixed.  So let us
construct new Fermi coordinates by a Lorentz transformation.  We
define
\be
x^{\alpha'} = \Lambda^{\alpha'}_\alpha x^\alpha\,,
\ee
where
$\Lambda$ is diagonal with
\blea
\Lambda^{u'}_u &=& \delta^{-1}\\
\Lambda^{v'}_v &=& \delta\\
\Lambda^{x'}_x &=& \Lambda^{y'}_y = 1\,.
\elea
In the primed coordinates, we have
\be\label{eqn:PhiVprime}
\Phi_V(\eta,\tau)=(\eta/\delta+ \tau/\sqrt{2}, \tau/\sqrt{2},0,0)\,.
\ee
Equation~(\ref{eqn:Rbound}) gives a bound on the components of the
Riemann tensor, measured in the original tetrad.  The covariant
components $R_{abcd}$ transform oppositely to the coordinate
components, so 
\be
R_{a'b'c'd'}=\Lambda^{a}_{a'}\Lambda^{b}_{b'}\Lambda^{c}_{c'}\Lambda^{d}_{d'}R_{abcd}\,,
\ee
where
\blea
\Lambda^u_{u'} &=& \delta \\
\Lambda^v_{v'} &=& \delta^{-1}\\
\Lambda^x_{x'} &=& \Lambda^y_{y'} = 1
\elea
Since we are taking $\delta \rightarrow \infty $, components of R with
more $u$'s than $v$'s diverge after the transformation. Components of R
with fewer $u$'s than $v$'s go to zero and components with equal
numbers of $u$'s and $v$'s remain the same.  We want the curvature to
be bounded by $\Rmax$ in the primed coordinate system, which will be
true if all components of the Riemann tensor with more $u$'s than
$v$'s are zero.  We will now show that this is the case in our system.

All points of interests are on achronal null geodesics, which thus
must be free of conjugate points.  Using
Eq.~(\ref{eqn:nullconvergence}) and proposition 4.4.5 of
Ref.~\cite{HawkingEllis}, each geodesic must violate the ``generic
condition''.  That is to say, we must have
\be\label{eqn:nongeneric}
\ell^c \ell^d \ell_{[a} R_{b]cd[e} \ell_{f]} = 0
\ee
everywhere in $M'$.

The only nonvanishing components of the metric in the tetrad basis
are $g_{uv} = g_{vu} = -1$ and $g_{xx} = g_{yy} = 1$.  The tangent
vector $\bl$ has only one nonvanishing component $\ell^u = 1$,
while the covector has only one nonvanishing component $\ell_v = -1$.
Thus Eq.~(\ref{eqn:nongeneric}) becomes
\be
\ell_{[a} R_{b]uu[e} \ell_{f]} = 0\,.
\ee
Let $j$, $k$, $l$, $m$ and $n$ denote indices chosen only from $\{x,y\}$.
Choosing $a=m$, $e=n$, and $a=f=v$ we find
\be\label{eqn:Ruumn}
R_{muun} = 0
\ee
for all $m$ and $n$.  Thus
\be\label{eqn:Ruu}
R_{uu} = 0\,.
\ee
Equation~(\ref{eqn:Ruu}) also follows immediately from the fact that
since $R_{uu}$ cannot be negative, any positive $R_{uu}$ would lead to
conjugate points.

If we apply the null convergence condition,
Eq.~(\ref{eqn:nullconvergence}), to $\mathbf{V}=\bE_{(u)}+
\epsilon \bE_{(m)}+ (\epsilon^2 /2) \bE_{(v)}$, where $\epsilon \ll 1$, we get
\be\label{eqn:NECepsilon}
R_{uu}+2R_{mu} \epsilon+O(\epsilon^2)\ge0\,.
\ee
Since $R_{uu} = 0$ from Eq.~(\ref{eqn:Ruu}), in order to have
Eq.~(\ref{eqn:NECepsilon}) hold for both signs of $\epsilon$, we must have
\be\label{eqn:Rmu}
R_{mu} = 0\,.
\ee
Since $R_{mu}=-R_{umvu}+g^{jk}R_{jmku}$,
\be\label{eqn:Ruuvm1}
R_{umvu} = g^{jk}R_{jmku}\,.
\ee

Now we use the Bianchi identity,
\be\label{eqn:Bianchi}
R_{luum;n}+R_{lunu;m}+R_{lumn;u}=0\,.
\ee
From Eq.~(\ref{eqn:Ruumn}), $R_{luum,n}=0$.  The correction to make
the derivatives covariant involves terms of the forms
$R_{auum}\nabla_n \bE^{(a)}_l$ and $R_{laum}\nabla_n \bE^{(a)}_u$.
Because of Eq.~(\ref{eqn:Ruumn}), the only contribution to the first
of these comes from $a=v$, which we can transform using
Eq.~(\ref{eqn:Ruuvm1}).  For the second, we observe that
$0=\nabla_n(\bE^{(v)}\cdot\bE^{(v)}) = 2
\nabla_n\bE^{(v)}\cdot\bE^{(v)} = 2\nabla_n\bE^{(v)}_u$, so $a=v$ does
not contribute.  Furthermore $R_{lumn;u} = R_{lumn,u}$, because the
$u$ direction is the single final direction in the coordinate
construction of Sec.~\ref{sec:coordinates}, and so in this direction
the tetrad vectors are just parallel transported.  Thus we find
\bea\label{eqn:Rulmndeq}
\frac{dR_{lumn}}{du} &=& g^{jk}[R_{jmku}\nabla_n \bE^{(v)}_l
  + R_{jlku}\nabla_n \bE^{(v)}_m
-R_{jnku}\nabla_m \bE^{(v)}_l
-R_{jlku}\nabla_m \bE^{(v)}_n]\\
&&+(R_{lkum}+ R_{lukm})\nabla_n \bE^{(k)}_u
+(R_{lknu}+ R_{lunk})\nabla_m \bE^{(k)}_u\,.\nonumber
\eea
Eq.~(\ref{eqn:Rulmndeq}) is a first-order differential equation in
the pair of independent Riemann tensor components $R_{xuxy}$ and
$R_{yuxy}$.  By assumption, the curvature and its derivative vanish
in the distant past, and therefore the correct solution to these
equations is
\be\label{eqn:Rulmn}
R_{lumn}= 0\,.
\ee
Eqs.~(\ref{eqn:Ruuvm1}) and (\ref{eqn:Rulmn}) then give
\be\label{eqn:Ruuvm}
R_{umvu} = 0\,.
\ee

Combining Eqs.~(\ref{eqn:Ruumn}), (\ref{eqn:Rulmn}), and
(\ref{eqn:Ruuvm}) and their transformations under the usual Riemann
tensor symmetries, we conclude that all components of the Riemann
tensor with more $u$'s than $v$'s vanish as desired.  It follows that
\be\label{eqn:Rboundprime}
|R_{a'b'c'd'}|<\Rmax\,.
\ee
everywhere in $M'$.

A similar argument using the Bianchi identity twice more would show that
$R_{abcd} = 0$ unless 2 of $a$, $b$, $c$, and $d$ are $v$, but we will
not need that result here.

\subsection{Timelike paths} \label{sec:timelike}
We would like to apply Eq.~(\ref{eqn:QNEI2}) to the paths in
Eq.~(\ref{eqn:PhiVprime}).  First we show that they are timelike.
Differentiating Eq.~(\ref{eqn:PhiVprime}), we find the components of the
tangent vector $\bk = d\Phi_V/d\tau$ in the primed Fermi coordinate
basis (not the tetrad basis),
\be\label{eqn:kprime}
k^{u'}=k^{v'}=\frac{1}{\sqrt{2}}\,.
\ee
The squared length of $\bk$ in terms of these components is
$g_{\alpha'\beta'}k^{\alpha'} k^{\beta'}$.  We showed in
Ref.~\cite{Kontou:2012kx} that $g_{\alpha'\beta}=
\eta_{\alpha'\beta'}+h_{\alpha'\beta'}$, where $h_{\alpha'\beta}$ at some
point $\bX$ is a sum of a small number of terms (6 in the present case
of 2-step Fermi coordinates) each of which is a coefficient no greater
than 1 times an average of
\be\label{eqn:RXX}
R_{\alpha'\gamma'\delta'\beta'}X^{\delta'}X^{\gamma'}
\ee
over one of the geodesics used in the construction of the Fermi
coordinate system.  The summations over $\delta'$ and $\gamma'$ in
Eq.~(\ref{eqn:RXX}) are only over restricted sets of indices depending
on the specific term under consideration.  From
Eqs.~(\ref{eqn:PhiVprime}) and (\ref{eqn:eta0}) the points under
consideration satisfy \bml\label{eqn:uvmax}
\bea
|u'|&<&u_1/\delta+\sqrt{2}\tau_0\label{eqn:umax}\\
|v'|&<&\tau_0/\sqrt{2}\\
x' &=& y' = 0\,.
\elea

From Eq.~(\ref{eqn:tau0}), the first term in Eq.~(\ref{eqn:umax})
decreases faster than the second, so we find that all components of
$\bX$ are $O(\tau_0)$.  Using Eq.~(\ref{eqn:Rboundprime}) we find
\be
h_{\alpha' \beta'} = O(\Rmax\tau_0^2)
\ee
so
\be \label{eqn:timelike}
g_{\alpha'\beta'}k^{\alpha'}k^{\beta'}=-1+O(\Rmax\tau_0^2)\,.
\ee
Thus for sufficiently large $\delta$, and thus small $\tau_0$, $\bk$
is timelike.

No we consider the acceleration of our paths.
Reference~\cite{Kontou:2012kx} gives the affine connection
$\nabla_{\beta'} E^{\gamma'}_{(\alpha')}$ as a sum of 2 averages of
terms of the form
\be 
{R^{\gamma'}}_{\alpha'\delta'\beta'} X^{\delta'}=O(\Rmax\tau_0)
\ee
just as above.  Thus the acceleration is given by
\be
|a^{\beta'}|=\frac{Dk^{\beta'}}{d\tau}
=|k^{\alpha'} \nabla_{\alpha'} k^{\beta'}|=|k^{\alpha'} k^{\gamma'} \nabla_{\alpha'} E_{(\gamma')}^{\beta'}| = O(\Rmax \tau_0)\,.
\ee
We want to show that the components of the acceleration are small, so we will calculate the dimensionless quantity
\be \label{eqn:propera}
|a^{\beta'}| \tau_0 = O(\Rmax \tau_0^2)\,.
\ee

\subsection{Causal diamond}
For each $\eta$, we would like to apply Eq.~(\ref{eqn:QNEI2}).  But
what is the spacetime $N$ in which we are to work?  It must
include the timelike path from $p = \Phi_V(\eta,-\tau_0)$ to
$q = \Phi_V(\eta,\tau_0)$, and to be globally hyperbolic it must
include all points in both the future of $p$ in the past of $q$, so we
can let $N$ be the ``double cone'' or ``causal diamond'',
\be
N=J^+(p) \cap J^-(q)\,.
\ee
We have shown that the curvature is small everywhere in the tube
$M'$, so we must show that $N \subset M'$.

From the previous section, we have that the metric in primed
coordinates can be written as
\be
g_{\alpha' \beta'}=\eta_{\alpha' \beta'}+h_{\alpha' \beta'}\,,
\ee
where $h_{\alpha' \beta'}$ consists of terms of the form
$R_{\alpha'\gamma'\delta'\beta'}X^{\delta'}X^{\gamma'}$.
The double cone in flat space obeys
\be\label{eqn:xyuvflat}
|x'|,|y'|,|v'|<\tau_0\,,
\ee
so the same is true at zeroth order in the Riemann tensor $R$.
Thus at zeroth order,
\be
h_{\alpha' \beta'} =O(\Rmax\tau_0^2)\,,
\ee
and so at first order in $R$,
\be
|x'|,|y'|,|v'|<\tau_0(1+O(\Rmax\tau_0^2))\,.
\ee
Since $\tau_0 \ll r$ for large $\delta$, we have
\be
|x'|,|y'|,|v'|<r\,.
\ee
Now we can replace the primed coordinates,
\blea
x'&=&x\\
y'&=&y\\
v'&=&v \delta\,,
\elea
so
\be
|x|,|y|,|v| < r\,,
\ee
and $N\subset M'$ as desired.

\subsection{Quantum Inequality}

We would now like to apply Eq.~(\ref{eqn:QNEI2}) to give a lower bound
on the integral of $T_{uu}$ on the paths $\Phi_V(\eta,\tau)$.  Because
of the ambiguity involving local curvature terms in Conjecture
1, we will first bound
\be
T'_{uu} = T_{uu} - C_{uu}\,,
\ee
where $C_{ab}$ the is the particular local curvature term for which
Conjecture 1 holds.  We will then show that $C_{uu}$ does not
contribute.

Equation~(\ref{eqn:Rboundprime}) shows that the curvature is small in
the tetrad basis transported according to the construction of
Sec.~\ref{sec:coordinates}.  These are not precisely the coordinates
used in the conditions of Conjecture 1, but the difference is of no
consequence, precisely because the curvature is small.
Equations~(\ref{eqn:timelike}) and (\ref{eqn:propera}) show that, for
sufficiently large $\delta$, $\Phi_V(\eta,\tau)$ is a timelike path
with small acceleration.  The parameter $\tau$ is not exactly the
proper time, but we show in Appendix A that this contributes only a
correction of order $\Rmax\tau_0^2$.  Thus Eq.~(\ref{eqn:QNEI2}) gives
\be\label{eqn:QNEI3}
\int_{-\tau_0}^{\tau_0}d\tau\, T'_{uu}(\Phi_V(\eta, \tau))f(\tau/\tau_0)^2 \geq \frac{(\ell_ak^a)^2}{12 \pi^2 \tau_0^4}\int_{-\tau_0}^{\tau_0}d\tau f''(\tau/\tau_0)^2 [1+c(\Rmax\tau_0^2)]\,,
\ee
where $c(\Rmax\tau_0^2)$ vanishes as $\tau_0\to 0$.
In the unprimed coordinates, the only nonvanishing covariant component
of $\bl$ is $\ell_v = -1$, so $\ell_a k^a = - k^v = -1/(\sqrt{2}\delta)$ so
\be
(\ell_ak^a)^2=\frac{1}{2\delta^2}\,.
\ee
Let
\be
F=\int f''(\alpha)^2 d\alpha=\frac{1}{\tau_0} \int f''(\tau/ \tau_0)^2 d\tau\,.
\ee
Then Eq.~(\ref{eqn:QNEI3}) becomes
\be
\int_{-\tau_0}^{\tau_0}d\tau\, T'_{uu}(\Phi_V(\eta, \tau))f(\tau/\tau_0)^2 \geq -\frac {F}{24 \pi^2 \delta^2 \tau_0^3}[1+c(\Rmax\tau_0^2)]\,.
\ee
Integrating in $\eta$ gives
\be \label{eqn:QNEI4}
\int_{-\eta_0}^{\eta_0}d\eta \int_{-\tau_0}^{\tau_0}d\tau\, T'_{uu}(\Phi_V(\eta, \tau))f(\tau/\tau_0)^2 \geq -\frac {F\eta_0}{12 \pi^2 \delta^2 \tau_0^3}[1+c(\Rmax\tau_0^2)]\,.
\ee
Now consider 
\be\label{eqn:Cuu}
\int_{-\eta_0}^{\eta_0}d\eta \int_{-\tau_0}^{\tau_0}d\tau\, C_{uu}(\Phi_V(\eta, \tau))\,.
\ee
Terms from Eqs.~(\ref{eqn:12H}) proportional to $g_{ab}$ do not
contribute, because $g_{uu}=0$.  Similarly $R_{uu}=0$ from
Eq.~(\ref{eqn:Ruu}). The term $R^{cd}R_{ucud}$ vanishes because
$R_{ucud} = 0$ unless $c = d = v$, from Eqs.~(\ref{eqn:Ruumn}) and
(\ref{eqn:Ruuvm}), while $R^{vv} = g^{uv} g^{uv} R_{uu} = 0$.

The remaining term is $R_{;uu}$.  As explained in conjunction with
Eq.~(\ref{eqn:Bianchi}), the covariant nature of the derivatives does
not matter, and $R_{;uu}$ is a total derivative in $u$.  In
Eq.~(\ref{eqn:Cuu}), it is integrated $d\eta$ which is just $du$.  In
the limit where $\eta_0 \to\infty$, the boundary term vanishes,
because the curvature is localized.  Thus $C_{uu}$ does not contribute
and we can use $T_{uu}$ in place of $T'_{uu}$ in
Eq.~(\ref{eqn:QNEI4}).

Now we compare Eq.~(\ref{eqn:QNEI4}) to Eq.~(\ref{eqn:etintegral}).
Equation~(\ref{eqn:QNEI4}) says that integral over the parallelogram
is no more negative that something that goes to zero in the
$\delta\to\infty$ limit as
\be
\frac{\eta_0} {\delta^2 \tau_0^3}\sim \delta^{2\alpha-1}\,.
\ee
Equation~(\ref{eqn:etintegral}) says that the same integral is more
negative than something that goes to zero as $\tau_0 \sim \delta^{-a}$.
Since $\alpha<1/3$, the lower bound in Eq.~(\ref{eqn:QNEI4})
goes to zero more quickly than the upper bound in
Eq.~(\ref{eqn:etintegral}).  Thus for sufficiently large $\delta$, the
lower bound will be above the upper bound, so they cannot
simultaneously be satisfied.  This contradiction proves Theorem 1.

\section{Discussion}\label{sec:discussion}

As discussed in Ref.~\cite{Graham:2007va}, to have an exotic spacetime
there would have to be violation of ANEC on achronal geodesics,
generated by a state of quantum fields in that same spacetime.  We
have proved, subject to Conjecture 1 and the various assumptions
above, that minimally-coupled, free quantum scalar fields can only
violate ANEC on geodesics traveling through parts of spacetime that
violate the null convergence condition.  Could it be that a single
effect both violates ANEC and produces the curvature that allows ANEC
to be violated?  The following heuristic argument casts doubt on this
possibility.

Suppose ANEC violation and NEC violation have the same source.  We
will say that they are produced by an exotic stress-energy tensor
$\Texotic$.  This $\Texotic$ gives rise to an exotic
Einstein curvature tensor,
\be\label{eqn:G}
\Gexotic = 8\pi\lpl^2 \Texotic\,,
\ee
in units where $c=\hbar=1$.  It is $\Gexotic$ that permits $\Texotic$
to arise from the quantum field.  Without $\Gexotic$, the spacetime
would obey the null convergence condition, and so, since $\Texotic$
violates ANEC, it would have to vanish.  A reasonable conjecture is
that as $\Gexotic \to 0$, $\Texotic \to 0$ at least
linearly.\footnote{Not, for example, changing discontinuously for
  infinitesimal but nonzero $\Gexotic$ or going as $\Gexotic^{1/2}$.}
Then we can write schematically
\be\label{eqn:Texoticbound}
|\Texotic| \lesssim  l^{-2} |\Gexotic|\,,
\ee
where $l$ is a constant length obeying $l\gg \lpl$.  The parameter
$l$, needed on dimensional grounds, might be the wavelength of some
excited modes of the quantum field.  Equation~(\ref{eqn:Texoticbound})
is just schematic because we have not said anything about the places
at which these tensors should be compared, or in what coordinate
system they should be measured.

Combining Eqs.~(\ref{eqn:G}) and (\ref{eqn:Texoticbound}), we find
\be
|\Texotic| \lesssim (\lpl/l)^2 |\Texotic|\,,
\ee
which is impossible since $l\gg \lpl$.

Given the assumptions of this paper, it appears that the only
remaining possibility for self-consistent achronal ANEC violation
using minimally coupled free fields is to have first a quantum field
that violates NEC but obeys ANEC, and then a second quantum field (or
a second, weaker effect produced by the same field) that violates ANEC
when propagating in the spacetime generated by the first field.  The
stress-energy tensor of the second field would be a small correction
to that of the first, but perhaps this correction might lead to ANEC
violation on geodesics that were achronal (and thus obeyed ANEC only
marginally) taking into account only the first field.  This idea seems
rather unlikely to us, and we will attempt to rule it out in future
work.

If one considers quantum scalar fields with non-minimal curvature
coupling, the situation is rather different.  Even classical
non-minimally coupled scalar fields can violate ANEC
\cite{Barcelo:1999hq,Barcelo:2000zf}, with large enough (Planck-scale)
field values.  However, as the field values increase toward such
levels, the effective Newton's constant first diverges and the becomes
negative.  Such situations may not be physically realizable.  If one
excludes such field values, some restrictions are known, but there are
no quantum inequalities of the usual sort
\cite{Fewster:2006ti,Fewster:2007ec}, and there are general
\cite{Visser:1994jb} and specific \cite{Urban:2009yt,Urban:2010vr}
cases where conformally coupled quantum scalar fields violate ANEC in
curved space.  It may be possible to control such situations by
considering only cases where a spacetime is produced self-consistently
by fields propagating in that spacetime, but the status of this
``self-consistent achronal ANEC'' for non-minimally coupled scalar
fields outside the large-field region is not known.

\appendix
\section{Proper time}

We start with $\Phi_V(\eta,\tau)$ given by  Eq.~(\ref{eqn:PhiV})
with tangent vector
\be\label{eqn:k}
k = \frac{\partial}{\partial \tau} \Phi_V(\eta,\tau)
= \left(\frac{\delta}{\sqrt{2}},\frac{1}{\sqrt{2\delta}}\right)
\ee
in the coordinate basis.  We would like to reparameterize the path
$\Phi_V(\eta,\tau)$ in terms of proper time, which we will denote
$\tau'$.  Then $g_{ab}k'^a k'^b=-1$ where $k'$ is the tangent vector
to the reparameterized path,
\be
k'=\frac{\partial}{\partial \tau'} \Phi_V(\eta,\tau(\tau'))
= \frac{k}{d\tau'/d\tau}
\ee
so
\be
\frac{d\tau'}{d\tau}=\sqrt{-g_{ab}k^ak^b} = \sqrt{h}
\ee
with
\be
h=1-h_{ab}k^a k^b\,.
\ee
Now the left hand side of Eq.~(\ref{eqn:QNEI3}) can be written
\bea
 \int_{-\tau_0}^{\tau_0} d\tau T'_{ab}(\Phi_V(\eta,\tau))k^a k^b
f(\tau/\tau_0)^2&=&\int_{-\tau_0}^{\tau_0} d\tau'
T'_{ab}(\Phi_V(\eta,\tau(\tau')))k'^a k'^b \sqrt{h} f(\tau/\tau_0)^2
\nonumber\\
&=&\int_{-\tau_0}^{\tau_0} d\tau' T'_{ab}(\Phi_V(\eta,\tau'))k'^a k'^b g(\tau')^2\,,
\eea
where we let
\be
g(\tau') \equiv f(\tau(\tau')/\tau_0)h^{1/4}\,.
\ee
Now we can apply the quantum inequality for the function $g$ and proper
time $\tau'$. Since the curvature and the function $f$ are
smooth, so is $g$.  We get
\be\label{eqn:QNEIA}
\int_{-\tau_0}^{\tau_0} d\tau' T'_{ab}(\Phi_V(\eta,\tau'))k'^a k'^b g(\tau')^2 \geq -\frac{(\ell_a k'^a)^2}{12 \pi} \int_{-\tau_0}^{\tau_0} d\tau' g''(\tau')^2[1+c(\Rmax \tau_0^2)]
\ee

Now let us determine $h$.  Since we are working only in the $u$-$v$
plane, we have two-step Fermi coordinates with one index in each step.
Thus we can use Eq.~(27) of Ref.~\cite{Kontou:2012kx} to get
\be
h_{ab}(\bX)=2F_{ab}=2\int_0^1 d\lambda \alpha_{2m}(\lambda)(1-\lambda)R_{acdb}(\bX_{(1)}+\lambda\bX_{(2)})X^d_{(2)}X^c_{(2)}
\ee
where $\bX_{(1)} = \Phi(0,v)$ and $\bX_{(2)} = \Phi(u,0)$.  Because of
the symmetry of the Riemann tensor, the only
nonvanishing case is
\be \label{eqn:hvv}
h_{vv}=2\int_0^1 d\lambda (1-\lambda)R_{vuuv}(\Phi(\lambda u, v))u^2
\ee
and
\be
h = 1-h_{vv}k^vk^v = 1-\frac{h_{vv}}{2\delta^2}
\ee
The maximum magnitude of $u$ is $u_1+\sqrt{2}\tau_0\delta$, so in the
limit $\delta\to\infty$, $h = 1+ O(\Rmax \tau_0^2)$.

We are not interested in $O(\Rmax\tau_0^2)$ correction terms, and we
will write $\approx$ to show that such terms have been ignored.  Thus
we can take $h \approx 1$, except where it is differentiated, and we
will not worry about the difference between $k'^a$ and $k^a$, and that
between $d\tau'$ and $d\tau$ on the right hand side of
Eq.~(\ref{eqn:QNEIA}).

We would like to write $g''$ in terms of $f''$ and a correction that
vanishes in the limit $\delta\to\infty$. So we will calculate the
derivatives of $g$,
\be
\frac{dg}{d\tau'}=\frac{1}{\sqrt{h}}\frac{dg}{d\tau}=
h^{-1/4}\frac{df}{d\tau}+\frac{f}{4}h^{-5/4}\frac{dh}{d\tau}
\approx \frac{df}{d\tau}+\frac{f}{4}\frac{dh}{d\tau}
\ee
\be \label{eqn:secg}
\frac{d^2g}{d\tau'^2}\approx
\frac{d^2f}{d\tau^2}-\frac{5f}{16}\left(\frac{dh}{d\tau}\right)^2
+\frac{f}{4}\frac{d^2h}{d\tau^2}
\ee
To compute the derivatives of $h$, we will change variables to
$q=\lambda u$ in Eq.~(\ref{eqn:hvv}) to get
\be
h=1-\frac{1}{\delta^2}\int_0^u dq (u-q)R_{vuuv}(\Phi(q,v))
\ee
Now we can calculate the first derivative,
\bea
\frac{dh}{d\tau}&=&k^u\frac{dh}{du}+k^v\frac{dh}{dv}
= \frac{\delta}{\sqrt{2}}\frac{dh}{du}
+\frac{1}{\sqrt{2}\delta}\frac{dh}{dv}\\
&=& -\frac{1}{\sqrt{2}\delta}\int_0^u dq\, R_{vuuv}(\Phi(q,v))
-\frac{1}{\sqrt{2}\delta^3}\int_0^u dq\, (u-q)R_{vuuv,v}(\Phi(q,v))\,.\nonumber
\eea
Using the bounds from Sec.~\ref{sec:curvature}, we find in the
$\delta\to\infty$ limit,
\be \label{eqn:firh}
\left | \frac{dh}{d\tau} \right | \leq \tau_0 \Rmax
+\frac{\tau_0^2}{\sqrt{2}\delta}\dRmax\,.
\ee
For sufficiently large $\delta$ the second term is negligible compared
to the first.

For the second derivative we can write
\bea 
\frac{d^2h}{d\tau^2}
&=&
\frac{\delta^2}{2}\frac{d^2h}{du^2}+\frac{d^2h}{dudv}+\frac{1}{2\delta^2}\frac{d^2h}{dv^2}\\
&=&-\frac{1}{2}R_{vuuv}-\frac{1}{\delta^2}\int_0^u dq\, R_{vuuv,v}(\Phi(q,v))
-\frac{1}{2\delta^4}\int_0^u dq\, (u-q)R_{vuuv,vv}(\Phi(q,v)))\nonumber
\eea
Again using the bounds from Sec.~\ref{sec:curvature}, we find
\be \label{eqn:sech}
\left |\frac{d^2h}{d\tau^2} \right | \leq
\frac{1}{2}\Rmax+\frac{\sqrt{2}\tau_0}{\delta}\dRmax
+\frac{\tau_0^2}{2\delta^2}\ddRmax
\ee
As before, for sufficiently large $\delta$, the second and third term
can be neglected in comparison to the first.

Keeping only the most important corrections, we then find
\be \label{eqn:app}
\left |\frac{d^2g}{d\tau'^2} \right | \lesssim
\frac{f''}{\tau_0^2} -\frac{5}{16}f\Rmax^2\tau_0^2
+\frac{1}{8}f \Rmax=\frac{1}{\tau_0^2}[f''+O(\Rmax\tau_0^2)]\,,
\ee
which justifies ignoring the difference between $\tau$ and $\tau'$ in
Eq.~(\ref{eqn:QNEI3}).

A similar argument applies to the acceleration.  In
Sec.~\ref{sec:timelike} we found that the acceleration was small,
\be\label{eqn:acopy}
\frac{Dk^{\beta'}}{d\tau} = O(\Rmax\tau_0)\,.
\ee
Changing to the proper time $\tau'$ means that we should consider
instead
\be
\frac{D{k'}^{\beta'}}{d\tau'} \approx \frac{D{k'}^{\beta'}}{d\tau} =
\frac{D}{d\tau}\left(\frac{k^\beta}{\sqrt{h}} \right)
\approx \frac{Dk^\beta}{d\tau}-\frac{1}{2}\frac{dh}{d\tau}k^{\beta'}
= O(\Rmax\tau_0)
\ee
from Eqs.~(\ref{eqn:acopy}), (\ref{eqn:firh}), and (\ref{eqn:kprime}).
Thus Eq.~(\ref{eqn:propera}) holds for the proper acceleration as
well.

\section*{Acknowledgments}

We would like to thank Chris Fewster, Larry Ford, Tom Roman, Ben
Shlaer, and Doug Urban for useful conversations.  This work was
supported in part by grants RFP2-06-23 and RFP3-1014 from The
Foundational Questions Institute (fqxi.org)

\bibliography{no-slac,paper}

\end{document}